\documentclass[aps,pre,one column,groupedaddress,11pt,notitlepage]{revtex4-1}
\usepackage{bm}
\newcommand{\ve}[1]{\mathbf{#1}}
\newcommand{\bmm}{\ve{m}}

\newcommand{\bu}{\ve{u}}

\newcommand{\bp}{\ve{p}}

\newcommand{\bq}{\ve{q}}

\newcommand{\bx}{\ve{x}}

\newcommand{\bs}{\ve{s}}

\newcommand{\bR}{\ve{R}}

\newcommand{\bF}{\ve{F}}
\newcommand{\bI}{\ve{I}}

\newcommand{\bphi}{\boldsymbol{\phi}}

\newcommand{\dd}{\text{d}}

\newcommand{\overbar}[1]{\mkern 1.5mu\overline{\mkern-1.5mu#1\mkern-1.5mu}\mkern 1.5mu}

\usepackage{epsf}
\usepackage{amsmath}
\usepackage{amssymb}
\usepackage{stmaryrd}
\usepackage{setspace}
\usepackage{enumitem}
\usepackage{graphicx}
\usepackage{float}
\usepackage{footnote}
\usepackage{microtype}
\usepackage{scalefnt}
\usepackage{microtype}
\usepackage{xfrac}
\usepackage{transparent}
\usepackage{rotate}
\usepackage{array}
\usepackage{tabu}
\usepackage{subcaption}
\usepackage[mediumspace,mediumqspace,Grey,squaren]{SIunits}
\usepackage{xcolor}

\begin{document}

\title{A quantitative comparison of phase-averaged models \\ for bubbly, cavitating flows}

\author{Spencer H. Bryngelson}
\email{Corresponding author: spencer@caltech.edu}
\author{Kevin Schmidmayer}
\author{Tim Colonius}
\affiliation{Division of Engineering and Applied Science,
California Institute of Technology,
1200 E California Blvd, Pasadena, CA 91125, USA}


\begin{abstract}
	We compare the computational performance of
	two modeling approaches for the flow of
	dilute cavitation bubbles in a liquid.
	The first approach is a deterministic model, for which bubbles are represented in a
	Lagrangian framework as advected features, each sampled from a distribution of
	equilibrium bubble sizes. The dynamic coupling to the liquid phase
	is modeled through local volume averaging.
	The second approach is stochastic; ensemble-phase
	averaging is used to derive mixture-averaged equations
	and field equations for the associated bubble properties are evolved in an Eulerian reference frame.
	For polydisperse mixtures, the probability density function of the equilibrium bubble radii
	is discretized and bubble properties are solved for each representative bin.
	In both cases, the equations are closed by solving Rayleigh--Plesset-like
	equations for the bubble dynamics as forced by the local or mixture-averaged
	pressure, respectively. An acoustically excited dilute bubble screen is used
	as a case study for comparisons. We show that observables of
	ensemble- and volume-averaged simulations match closely and that
	their convergence is first order under grid refinement.  Guidelines are established
	for phase-averaged simulations by comparing the computational costs of methods.
	The primary costs are shown to be associated with stochastic closure;
	polydisperse ensemble-averaging requires many samples of the underlying
	PDF and volume-averaging requires repeated, randomized simulations to accurately
	represent a homogeneous bubble population. The relative sensitivities
	of these costs to spatial resolution and bubble void fraction are presented.
\end{abstract}

\maketitle

\section{Introduction}\label{s:intro}

We consider the modeling of a flowing mixture of dilute cavitation bubbles.
The associated phenomenologies are
often complex: bubbles can oscillate, expand significantly (cavitate), and collapse
violently. Notably, the multiphase bulk flow is
sensitive to individual bubble motion; the shock-waves emitting from a
cavitation event are often comparable to those in the bulk
flow~\citep{reisman98,brennen95}, and even just a few bubbles are sufficient to
modify larger-scale pressure waves~\citep{mettin03}. While the flow of dilute, cavitating bubbles 
is only one possible bubbly flow scenario in a host of others, such as buoyancy-driven bubble 
motion~\citep{risso18}, the broad natural occurrence
and application of this subset motivates further study of their behavior.
Such bubbles emerge naturally via, e.g., cavitation nucleation in
mantis~\citep{patek04,patek05} and pistol shrimp
strikes~\citep{bauer04,koukouvinis17}, bubble-net feeding of humpback
whales~\citep{leighton04_2,leighton07_2}, and vascular plant
tissues~\citep{pickard81}. Dilute bubbly suspensions are also generated in engineering flow applications;
purposefully, bubbles are nucleated in biomedical settings, including shock wave
lithotripsy~\citep{coleman87,pishchalnikov03,ikeda06}, as shielding via
bubble screens~\citep{surov99}, and during underwater
explosions~\citep{etter13,kedrinskii76}. Unfortunately, cavitation is also an undesirable
consequence of the flow dynamics in other applications. For example, cavitation causes 
erosion, noise, and performance loss of pipe systems~\citep{weyler71,streeter83}, hydraulic
machinery~\citep{brennen95,naude61}, and propellers~\citep{sharma90,ji12}.

A theoretical understanding of complex bubbly flows is often prohibitive without
significant simplifications. Further, a vast range of scales is usually present.
The radius of single bubbles can be as small as microns and can grow to
as large as millimeters~\citep{brennen95}, whereas bubble clouds and turbulent
features are often on the order of meters or larger~\citep{d83}; the natural
frequency and nominal collapse times are usually on the order of microseconds,
and the flow observation time scale is on the order of
seconds~\citep{brennen95}. This makes computer simulations of the fully-resolved
flow dynamics prohibitive. Instead, modeling techniques are required
to accurately represent the flow.

The first models for dilute bubbly flows include theories for linear
scattering~\citep{foldy45} and nonlinear oscillatory
systems~\citep{iordanskii60,kogarko64,van64,van84}. Since then, most models can
be broadly classified as either ensemble-~\citep{zhang94} or
volume-averaging~\citep{commander89}. Herein, we focus on two specific examples,
one of each model, and assess their relative computational cost and convergence.

We first discuss the technical differences between
ensemble- and volume-averaged models in section~\ref{s:model_comparison}.
The mathematical formulation
of each method is presented in section~\ref{s:model} and the numerical methods used
to solve the associated equations are outlined in section~\ref{s:numerics}.
In section~\ref{s:results}, we consider an acoustically excited bubble screen and 
compare the computational costs and convergence of the methods. 
Key points and conclusions are discussed in section~\ref{s:conclusions}.

\section{Outset model comparison}\label{s:model_comparison}

\begin{figure}
	\centering
	\includegraphics[scale=1]{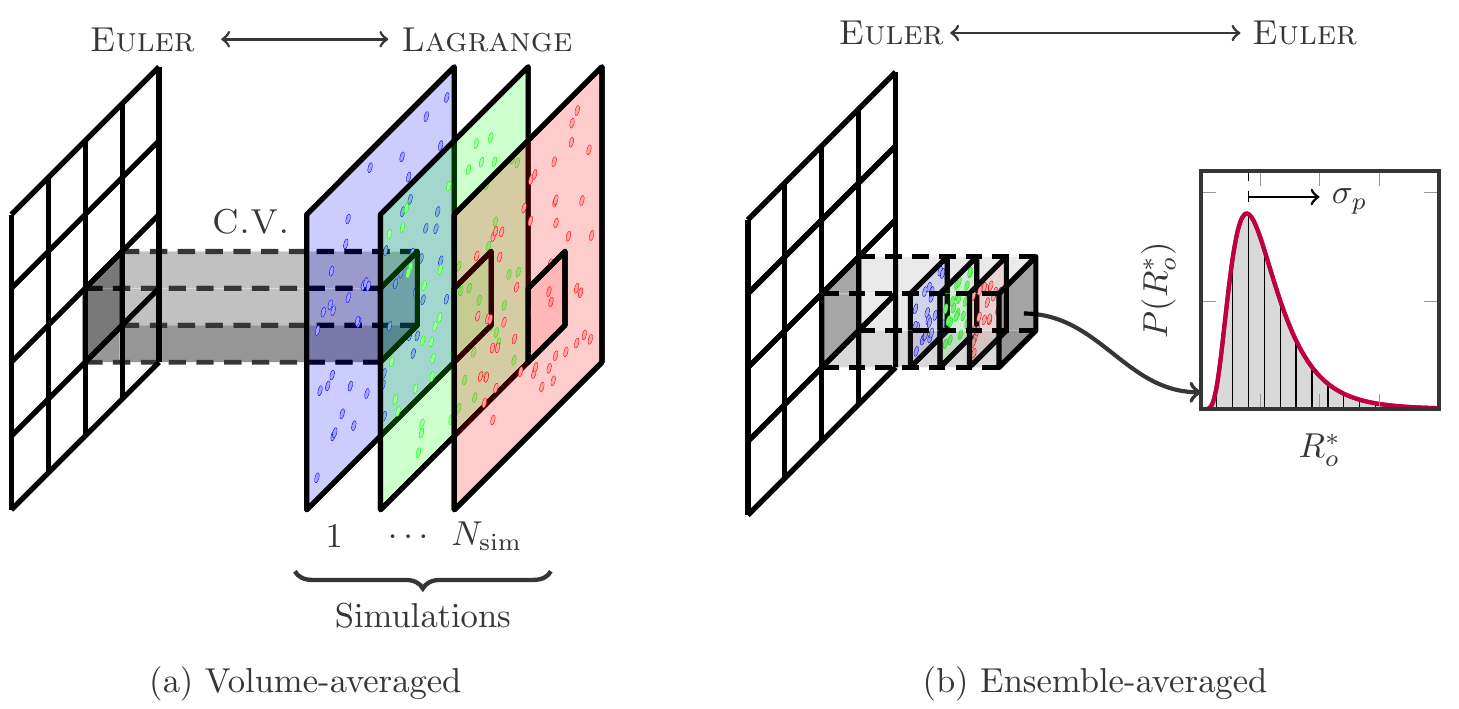}
	\caption{Schematic of (a) volume- and (b) ensemble-averaging models.
	}
	\label{f:models}
\end{figure}

The mixture-averaged flow equations associated with both ensemble- and volume-averaged
techniques represent the bubbles as features that interact with the flow. However, the bubbles are
tracked and coupled to the liquid phase differently. 
Volume-averaged models are formulated in an Euler--Lagrange framework, where individual bubbles are
Lagrangian particles, each 
sampled from an underlying spatial distribution (see figure~\ref{f:models}~(a)).
The volume of gas per-unit-volume of the mixture is obtained
locally for each computational cell by projecting the volume of bubbles onto the
grid. The disturbances induced by the bubbles on
the flow is computed by decomposing the potential generated inside each
cell into background and bubble parts: the background flow is constant inside a
cell, whereas the potential generated by each bubble decays with the distance
from the bubble center~\citep{fuster11}. The ensemble-averaged approach is an Euler--
Euler method and is  depicted in figure~\ref{f:models}~(b); instead of solving for the dynamics 
of individual bubbles, it evaluates the statistically-averaged 
mixture dynamics by assuming a large number of stochastically scattered bubbles dispersed within 
each computational grid cell~\citep{ando11}.

Besides algorithmic differences, there are also differences in assumptions that lead
to their respective closures. In the volume-averaged case, for the mixture dynamics to 
be considered homogeneous, the length scale of the averaging volume (shown in figure~\ref{f:models}~(a)) must be much
larger than mean bubble spacing and much smaller than the mixture length scale~\citep{nigmatulin79,prosperetti01}. 
Ensemble-averaged models are not beholden to this assumption from the outset, though 
ultimately the separation of scales is still invoked for model closure.
In these theoretical limits, ensemble- and volume-averaging are statistically equivalent 
procedures~\citep{batchelor70,biesheuvel84}; however, neither the sensitivity of their 
respective closures to under-resolution nor their computational costs have 
been compared for practical simulations. Herein, we formally contrast these observables.

\section{Mathematical model formulation}\label{s:model}

We describe the flow of a dilute suspension of dynamically evolving bubbles in a compressible liquid.
Ensemble- and volume-averaged models are presented; in addition to the assumptions of
section~\ref{s:model_comparison}, we assume that there is no-slip between the gas and liquid phases
and that the gas density is much smaller than the liquid density. While phase-slip is required 
to describe key phenomenologies for some bubbly flows, such as buoyancy-driven mixing 
flows~\citep{risso18}, they are thought to play a lesser role in the cavitating bubble dynamics we consider 
here~\citep{matsumoto96}. The average mixture equations of motion
take their usual quasi-conservative form~\citep{commander89}:
\begin{gather}
	\frac{ \partial \bq }{ \partial t } + \nabla \cdot \bF = \bs 
	\label{e:goveq}
\end{gather}
where $\bq = \left\{ \rho, \rho \bu, E \right\}$ are the conservative variables, $\bF = 
\left\{ \rho \bu, \rho \bu \bu + p \bI, (E + p) \bu \right\}$ are the fluxes, and $\bs$ 
are the source terms associated with bubble modeling.
Here, $\rho$, $\bu$, $p$, and $E$ are the mixture density, velocity vector, pressure,
and total energy, respectively. Mixture variables obey $(\cdot) = (1-\alpha)(\cdot)_l + \alpha (\cdot)_g$,
where $\alpha$ is the void fraction and $l$ and $g$ denote the liquid and gas states, respectively.

\subsection{Ensemble-averaged flow equations}

Our formulation of the continuum ensemble-averaged equations
generally follows that of \citet{zhang94}.
The equilibrium radii of the bubble population 
are represented discretely as $\bR_o$, which are $N_\text{bin}$ bins of an assumed
log-normal PDF with standard deviation $\sigma_p$~\citep{colonius08}. 
The bins are distributed using a Gauss--Hermitian quadrature,
though previous works have shown that similar results are obtained with the same number
of quadrature points when using Simpson's rule~\citep{ando10}.
The instantaneous bubble radii are a function of these equilibrium states as 
$\bR(\bR_o) = \{ R_1, R_2, \dots, R_{N_\text{bin}} \}$.
In this case, $\bs = \mathbf{0}$. The mixture pressure is
\begin{gather}
	p = (1-\alpha)p_l +
	\alpha  \left(
		\frac{\overbar{\bR^3 \bp_{bw} }}{\overbar{ \bR^3}} - \rho \frac{ \overbar{ \bR^3 \dot{\bR}^2 }}{ \overbar{\bR^3} }
	\right),
\end{gather}
where $p_l$ is the liquid pressure, which we model using the stiffened-gas equation of state \citep{menikoff89}
\begin{gather}
	\Gamma_l p_l = \frac{1}{1-\alpha} \left( E - \frac{1}{2} \rho v^2 \right) - \Pi_{\infty,l},
	\label{e:SEOS}
\end{gather}
$\dot{\bR}$ are the bubble radial velocities, and $\bp_{bw}$ are the bubble wall pressures.
The equation of state is parameterized by the specific heat ratio $\gamma_l  = 1 + 1/\Gamma_l$
and $\Pi_\infty$ is the liquid stiffness.
Overbars $\overbar\cdot$ denote the usual moments with respect to the log-normal PDF. 
The void fraction is transported as
\begin{gather}
	\mathrm{D}_t \alpha =
	3 \alpha \frac{ \overbar{\bR^2 \dot{\bR} }}{ \overbar{\bR^3} },
\end{gather}
where $\mathrm{D}_t \equiv \partial_t + \bu \cdot \nabla$ is the substantial derivative operator.
The bubble dynamics are evaluated as
\begin{gather}
	\frac{ \partial n \bphi}{\partial t} + \nabla \cdot (n \bphi \bu) = n \dot\bphi,
\end{gather}
where $\bphi \equiv \left\{ \bR, \dot\bR, \bp_b, \bmm_v \right\}$ is the vector of 
bubble dynamic variables ($\bp_b$ is the bubble pressure and $\bmm_v$ is the vapor mass;
model details are described in section~\ref{s:bubble}) and $n$ is the conserved 
bubble number density per unit volume
\begin{gather}
	n = \frac{3}{ 4 \pi} \frac{\alpha}{ \overbar{\bR^3} }.
\end{gather}

\subsection{Volume-averaged flow equations}\label{s:volavg}

Our volume-averaged approach follows that of~\citet{maeda18}. 
There are $N_\text{bub}$ bubbles, each located at $\bx_i$, $i = 1,2,\dots,N_\text{bub}$
and tracked as a Lagrangian point. The continuous void fraction field $\alpha(\bx)$ is defined 
via volumetric bubble smearing as
\begin{gather}
	\alpha(\bx) = \sum_{i=1}^{N_\text{bub}} V_i \, \delta(d_i,h),
\end{gather}
where $\delta$ is the Gaussian regularization kernel,
\begin{gather}
	 \delta(d_i,h) =
	 \begin{cases}
	 		\frac {1}{ (2\pi)^{3/2} h^3 } e^{ - \frac{d_i^2 } { 2 h^2 } },  &  0 \leq d_i < 3h, \\
	 		0, & 3h \leq d_i,
		\end{cases}
\end{gather}
$V_i$ is the bubble volume, $d_i = |\bx - \bx_i|$, and $h$ is the kernel support width.
The void fraction advects as
\begin{gather}
	\frac{ \partial \alpha(\bx) }{ \partial t } = \frac{ \partial }{ \partial t } \sum_{i = 1}^{N_\text{bub}} V_i \delta =
	\sum_{i = 1}^{N_\text{bub}} \frac{ \partial V_i }{ \partial t } \delta + \sum_{i = 1}^{N_\text{bub}} V_i \frac { \partial \delta } { \partial t },
\end{gather}
where
\begin{gather}
	\frac{ \partial V_i }{ \partial t } = 4 \pi R_i^2 \dot{R}_i
	\quad \text{and} \quad
	\frac{ \partial \delta }{ \partial t } = - \bu \cdot \nabla \delta.
\end{gather}
Thus, the source terms of~\eqref{e:goveq} for the volume-averaged method are
\begin{gather}
	\bs = \frac{\bq}{1-\alpha} \mathrm{D}_t \alpha,
\end{gather}
which transport the void fraction on the mesh, and the 
mixture pressure is simply $p = p_l(1-\alpha)$, for which 
$p_l$ is recovered from \eqref{e:SEOS}.

\subsection{Single-bubble dynamics}\label{s:bubble}

We model the single-bubble dynamics under the assumption that the 
bubbles remain a spherical, ideal, and spatially uniform gas region, which does 
not interact with other bubbles, break-up, or coalesce.
The bubble dynamics are driven by pressure fluctuations of the surrounding liquid;
in our model, their radial accelerations $\ddot{R}$ are computed by the Keller--Miksis equation~\citep{keller80}:
\begin{gather}
	R \ddot{R} \left( 1 - \frac{ \dot{R} }{c} \right)   + \frac{3}{2} \dot{R}^2 \left( 1 - \frac{ \dot{R} }{3c} \right) =
	\frac{ p_{bw} - p_\infty }{\rho} \left( 1 + \frac{ \dot{R} }{c} \right) + \frac{ R \dot{p}_{bw} }{\rho c},
\end{gather}
where $c$ is the speed of sound, $p_\infty$ is the bubble forcing pressure, and
\begin{gather}
	p_{bw} = p_b - \frac { 4 \mu \dot{R} } { R } - \frac { 2 \sigma } { R }
\end{gather}
is the bubble wall pressure, for which $p_{b}$ is the internal bubble pressure, $\sigma$ is the surface tension coefficient,
and $\mu$ is the liquid viscosity. The evolution of $p_b$ is evaluated using the model of~\citet{ando10}:
\begin{gather}
	\dot{p}_{b} = \frac { 3 \gamma _ {b} } {R} \left( - \dot{R} p_{b} + \mathfrak{R}_{v} T_{bw}
	\dot{m}_v + \frac{ \gamma_b - 1 } { \gamma_b } k_{bw}  \left.  \frac { \partial T } { \partial r } \right|_{r=R}  \right),
\end{gather}
where $T$ is the temperature, $k$ is the thermal conductivity,
$\mathfrak{R}_v$ is the gas constant, $\gamma_b$ is the specific heat ratio of the gas,
and subscript $w$ indicates properties evaluated at the bubble wall $r = R$.
Mass transfer of the bubble contents follows the reduced model of~\citet{preston07}:
\begin{gather}
		\dot{m}_v = \frac { \mathcal{D} \rho_{bw} }{ 1 - \chi_{vw} } \left. \frac { \partial \chi_{v} } { \partial r } \right|_{r=R},
\end{gather}
where $\chi_v$ is the vapor mass fraction and $\mathcal{D}$ is the binary diffusion coefficient.
This single-bubble model includes thermal effects, viscous and acoustic damping, and phase change,
and its full formulation and ability to represent actual bubble dynamics have been presented 
elsewhere~\citep{preston07,ando10}.

\section{Numerical methods}\label{s:numerics}

Our numerical scheme generally follows that of \citet{coralic14}.
For this, the spatial discretization of \eqref{e:goveq} 
in three-dimensional Cartesian coordinates is
\begin{gather}
	\frac{ \partial \bq }{\partial t} +
	\frac{\bF^x(\bq)}{\partial x} +
	\frac{\bF^y(\bq)}{\partial y} +
	\frac{\bF^z(\bq)}{\partial z} =
	\bs(\bq),
	\label{e:partial}
\end{gather}
where $\bF^i$ are the $i \in (x,y,z)$ flux vectors. We
spatially integrate \eqref{e:partial} within each cell-centered
finite volume as
\begin{gather}
	\frac{ \dd \bq_{i,j,k} }{\dd t} +
	\frac{1}{\Delta x_i} [\bF_{i+1/2,j,k}^x - \bF_{i-1/2,j,k}^x ] +
	\frac{1}{\Delta y_j} [\bF_{i,j+1/2,k}^y - \bF_{i,j-1/2,k}^y ] +
	\frac{1}{\Delta z_k} [\bF_{i,j,k+1/2}^z - \bF_{i,j,k-1/2}^z ] =
	\bs(\bq_{i,j,k}).
\end{gather}
We reconstruct the primitive variables at the finite-volume-cell faces via a 5th-order WENO scheme~\citep{coralic14}
and use the HLLC approximate Riemann solver to compute the fluxes~\citep{toro94}. The time derivative
is computed using the 3rd-order TVD Runge--Kutta algorithm~\citep{gottlieb98}.

\section{Results}\label{s:results}

\subsection{Dilute bubble screen setup}

We consider an acoustically excited dilute bubble screen as a case study of the
differences and behaviors of the ensemble- and volume-averaged flow
models. Indeed, bubble screens are a practical configuration;
they can serve as a reduced model for dilute bubble clouds
and have been utilized to mediate structural damage due to underwater 
explosions~\citep{langefors67,domenico82},
modify the open-channel flow topology~\citep{blackaert08}, and even manipulate 
fish behavior~\citep{patrick85}.

\begin{figure}
	\centering
	\includegraphics[scale=1]{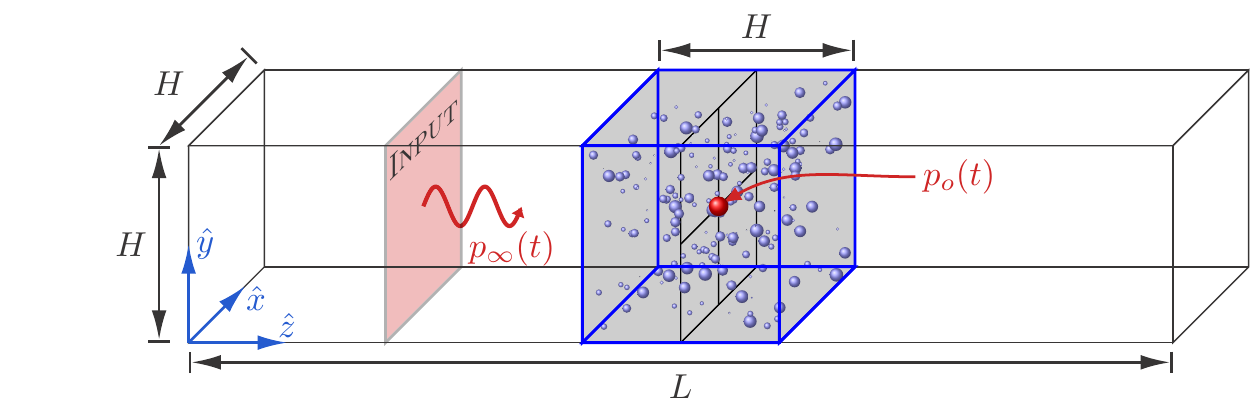}
	\caption{The model flow system.}
	\label{f:model}
\end{figure}

A schematic of the problem setup is shown in figure~\ref{f:model}. The
domain is a square prism with $x,y \in [-H/2,H/2]$ and $z \in [ -L/2,L/2 ]$,
where $L = \unit{25}{\milli\meter}$ and $H = L/5$; the boundaries are
non-reflective. A bubble screen occupies the cubic region $x,y,z \in
[-H/2,H/2]$, wherein the initial void fraction is $\alpha_o = 4 \times 10^{-5}$.  
In the volume-averaged case, the bubble positions are distributed uniformly in 
the bubble-screen region via a pseudo-random number generator. The bubbles
are initially quiescent with mean (and equilibrium) radii uniformly sampled from a log-normal
distribution centered at $R_o^\ast = \unit{10}{\micro\meter}$ with standard deviation $\sigma_p$.
For the volume-averaged method, this distribution must be sampled 
multiple times via independent simulations to represent the homogeneous mean flow.
Ensemble-averaging instead samples the most-probable equilibrium bubble radii 
$N_\text{bin}$ times and solves the corresponding bubble dynamic equations for the 
statistically-homogeneous flow; the relative costs of these
procedures are examined in section~\ref{s:closure}.
The Cartesian grid has uniform mesh spacing with $N_z =
250$ and $N_x = N_y = 50$ unless otherwise stated. 
The initial condition is quiescent at ambient pressure. A plane acoustic source at $z =
-3H/2$ excites one cycle of a $\unit{300}{\kilo\hertz}$, $p_A =
\unit{100}{\kilo\pascal}$ sinusoidal pressure wave $p_\infty$ in the $+z$
direction. The liquid is water with specific heat ratio $\Gamma = 0.16$ and
stiffness $\Pi_\infty = \unit{356}{\mega\pascal}$~\citep{maeda18}.

\subsection{Comparison of observables}

\begin{figure} 
	\centering
	\includegraphics[scale=1]{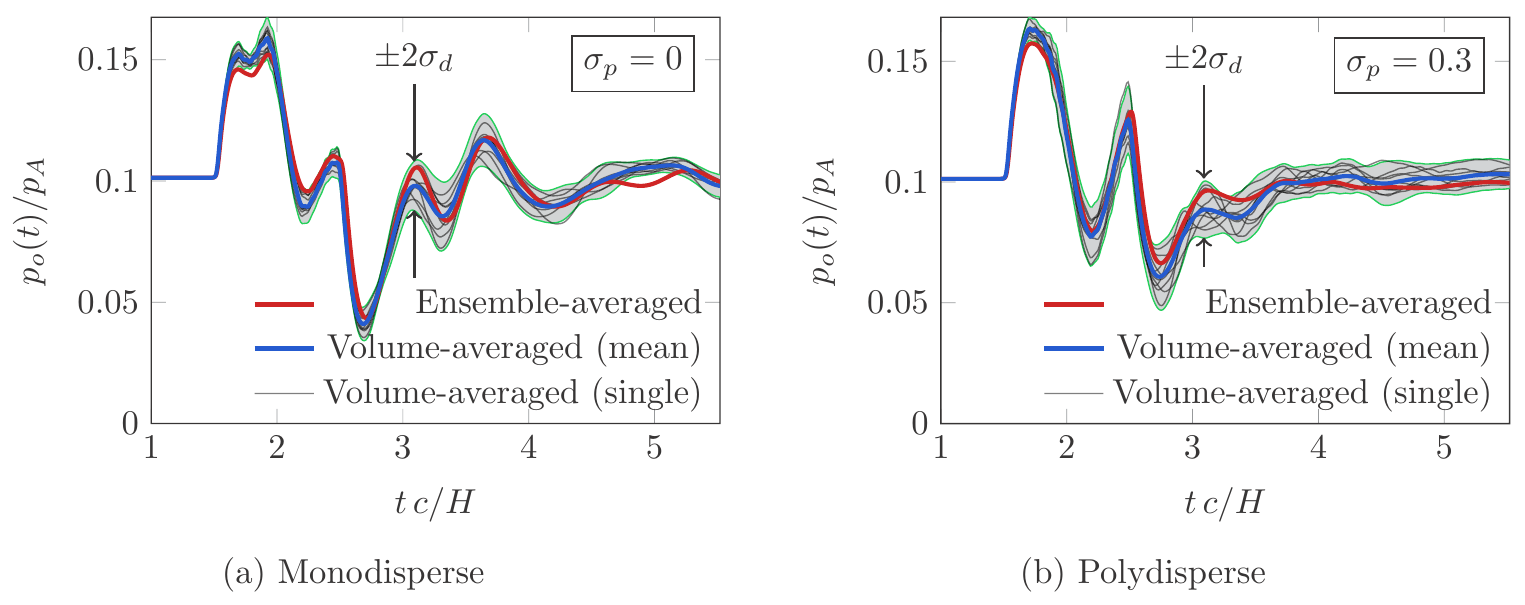}
	\caption{
	Pressure $p_o$ for (a) monodisperse ($\sigma_p = 0$) and (b) polydisperse ($\sigma_p = 0.3$) bubble distributions. 
	Individual volume-averaged simulations, each initialized with a different randomized
	bubble population, and their mean are shown, as well as the ensemble-averaged
	result.
	}
	\label{f:probe}
\end{figure}

We compare the mixture-averaged pressure at the bubble-screen center, $p_o
\equiv p(\{x,y,z\}=\{0,0,0\})$, for both methods.
Figure~\ref{f:probe} shows $p_o$ for (a) monodisperse and
(b) polydisperse bubble screens; the pressure grows then decays with the passage 
of the wave, with additional oscillatory features present due to the trapping of the wave 
in the screen region. In the volume-averaged case, multiple
simulations are averaged to determine the homogeneous
statistics; in figure~\ref{f:probe} these are labeled as ``single'' and ``mean'',
respectively. We compute the volume-averaged mean from 50 such
simulations, which have standard deviation $\sigma_d$.  In both cases, the
volume-averaged mean and ensemble-averaged pressures match closely, with the
difference within $2 \sigma_d$ of the individual volume-averaged pressures
almost everywhere.

\subsection{Spatial convergence}

\begin{figure}
	\centering
	\includegraphics[scale=1]{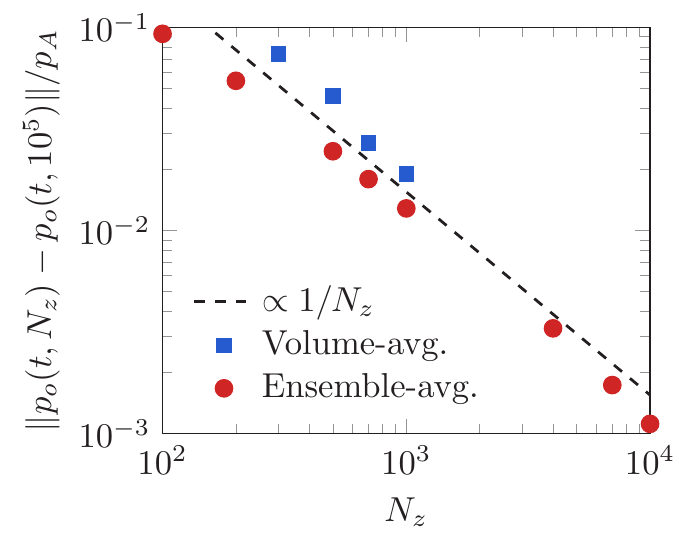}
	\caption{
	Spatial convergence of bubble-screen-centered mixture pressure.
	}
	\label{f:spatial_conv}
\end{figure}

We next evaluate the simulation response to spatial under-resolution. From the outset, it
is unclear if the response should be regular, as sub-grid modeling
can introduce mesh-dependent features. Figure~\ref{f:spatial_conv} shows the
$L_2$ difference between the bubble-screen centered mixture pressure and a
high-resolution simulation for both methods.
In the volume-averaged case, the mean pressure is used for comparisons and is computed via 200 individual 
simulations at each spatial resolution. This, coupled with the necessity of 3D 
simulations in the volume-averaged case (due to spatial heterogeneity), 
limits the largest $N_z$ we consider. 
We see that both methods monotonically and linearly converge.
Indeed, only linearity can be expected despite higher-order numerical methods,    
owing to the material discontinuities present and the truncation errors 
accumulated in the model assumptions, and thus stochastic closures. 
While the relative error we show here is smaller for the ensemble-averaged simulations,
we emphasize that the methods we use to compute it preclude a direct comparison of their 
accuracy. Furthermore, the cost of considering a small number of additional 
spatial mesh points is relatively small compared to that of the 
stochastic closures discussed next.

\subsection{Achieving stochastic closure and convergence}\label{s:closure}

\begin{figure} 
	\centering
	\includegraphics[scale=1]{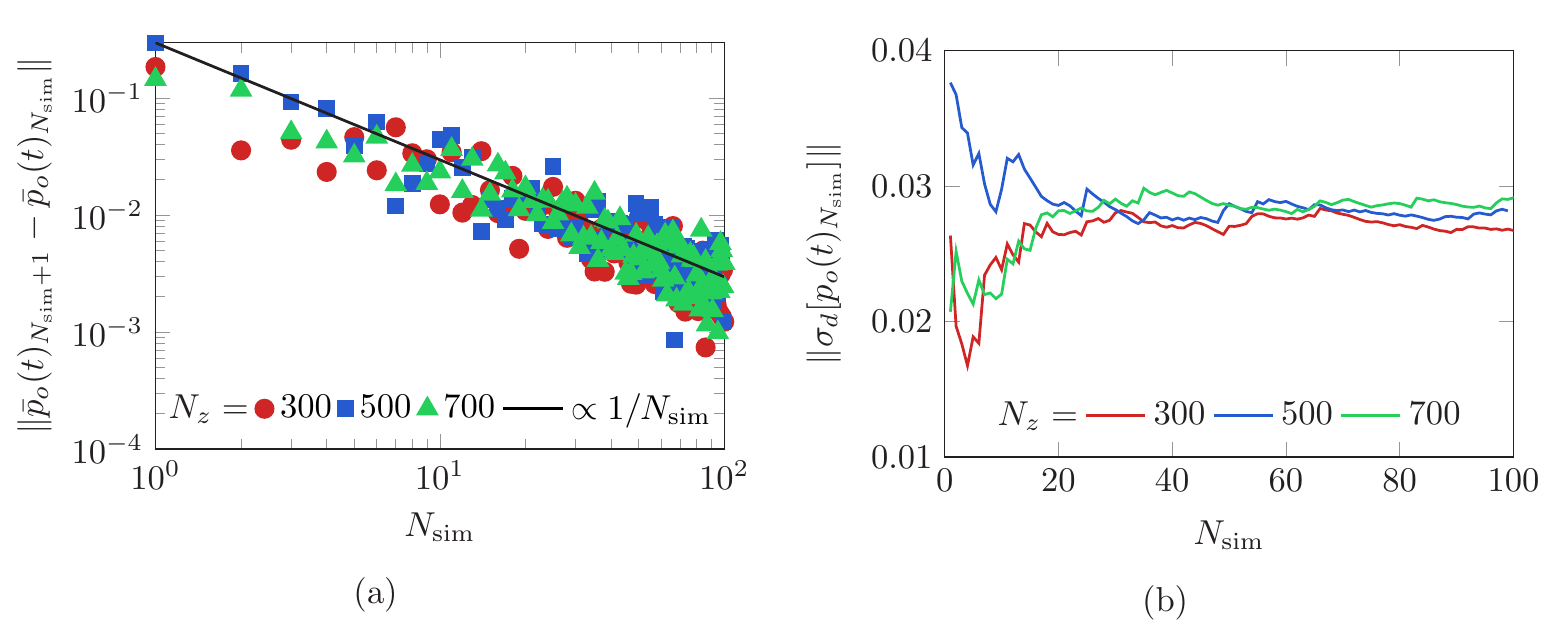}
	\caption{ 
	Convergence of bubble-screen-centered statistics with additional simulations $N_\text{sim}$;
	(a) mean pressure and (b) standard deviation 
	}
	\label{f:volavg_closure}
\end{figure}

Following the previous discussion, in the volume-averaged case 
multiple volume-averaged simulations, each initialized with a different
sample of the bubble size and position distributions, are
required to converge to the homogeneous mean flow. 
Figure~\ref{f:volavg_closure}~(a) shows the mean bubble-screen-centered pressure $\bar{p}_o$ 
for additional simulations $N_\text{sim}$; we see that it converges as $1/N_\text{sim}$ for all $N_\text{sim}$ and $N_z$.
The value of the difference is important when considering how many simulations
are required for a given accuracy.

Figure~\ref{f:volavg_closure}~(b) shows the standard deviation of the
individual simulations associated with $\bar{p}_o$, $\lVert
\sigma[p_o(t)_{N_\text{sim}}]$. Its value is transient for $N_\text{sim} \lesssim 40$ and all
$N_z$, and is relatively constant for $N_\text{sim} \gtrsim 40$. That is, at
least 40 simulations are required for a faithful estimation of $\sigma_d$.
Generally, $\lVert \sigma_d [p_o(t)] \rVert$ increases with increasing
$N_z$, though this change is relatively small.

\begin{figure}[b]
	\centering
	\includegraphics[scale=1]{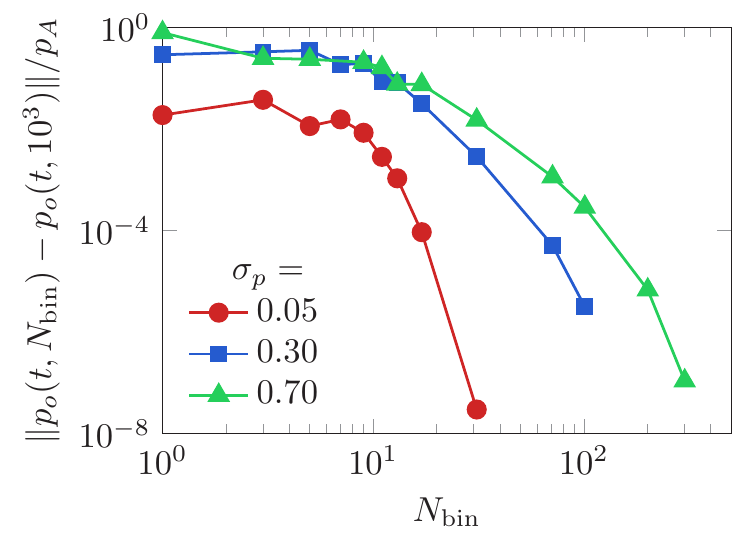}
	\caption{
	Convergence of ensemble-averaged simulations with $N_\text{bin}$ for
	varying degrees of polydispersity $\sigma_p$.
	}
	\label{f:ensavg_closure}
\end{figure}

For polydisperse ensemble-averaged simulations, the bubble size distribution,
given by a log-normal PDF, must be sampled multiple ($N_\text{bin}$) times. Sampling this
distribution is expensive, as each sample adds four equations and variables for each grid cell.
In figure~\ref{f:ensavg_closure}, we show the convergence of bubble-screen-centered
pressure with $N_\text{bin}$, as compared with a well-resolved $N_\text{bin} = 10^3$ simulation.
Convergence appears to be exponential, with generally larger error for 
larger $\sigma_p$ and fixed $N_\text{bin}$. This is expected, as larger $\sigma_p$ 
represents a broader bubble size distribution and thus, more samples are required
for the same accuracy. Small $N_\text{bin}$ entails relatively large error; for $\sigma_p = 0.3$, 
$N_\text{bin} = 11$ gives an error of
$8\%$ of $p_A$, whereas $N_\text{bin} = 101$ gives an error of only $10^{-4} \%$. 
We thus anticipate that, in this case, greater than $N_\text{bin} = 11$, but
less than $N_\text{bin} = 101$ samples are sufficient for most purposes.

\subsection{Computation cost}

We compute the computational cost of each method by considering the time-step
cost of a simulation configuration.  
For this benchmarking, simulations were performed
with the same three-dimensional grid and matching 
time-step sizes on a single core of a twelve-core Intel Xeon E5-2670 Haswell
$\unit{2.3}{\giga\hertz}$ processor. 
Here, $T_s$ is the time-step costs in seconds, which is
computed as the
average cost of a time-step over 1000 time steps of a single simulation.
We emphasize that both methods have the same simulation platform, following the
general implementation of~\citet{coralic14}, which ensures that the relative costs
computed for each method are restricted to the computational bubble model itself.

\begin{figure} 
	\centering
	\includegraphics[scale=1]{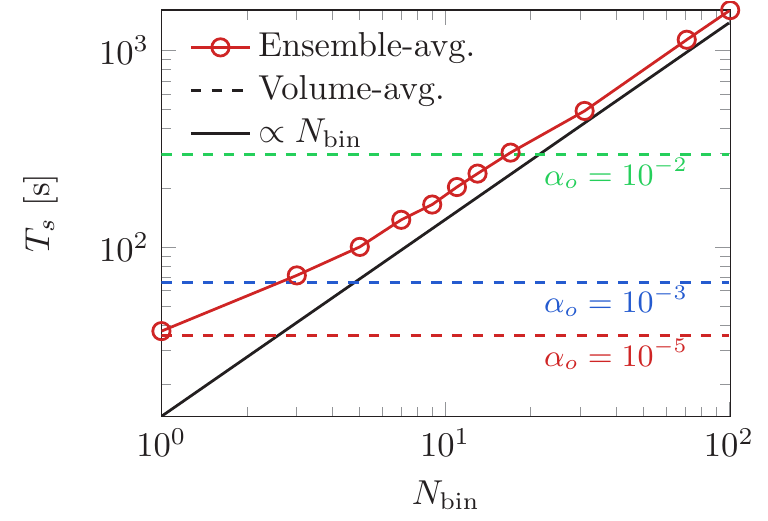}
	\caption{
	Time-step cost $T_s$ for simulations of varying
	polydispersity resolution $N_\text{bin}$ and initial void fraction $\alpha_o$.
	}
	\label{f:cost_nb}
\end{figure}

Figure~\ref{f:cost_nb} shows the relative
cost of polydisperse simulations. Polydisperse volume-averaged simulations are
no more costly than monodisperse simulations, so $T_s$ is independent of 
$N_\text{bin}$. The dashed lines show their cost for different initial void fractions, $\alpha_o$,
vary because larger $\alpha_o$ entails more bubbles, each of which is
evolved via the Keller--Miksis equation. For the ensemble-averaged simulations,
$T_s$ is independent of $\alpha_o$, since it is represented as an
Eulerian variable on the mesh. Instead, following the previous subsection, the cost
is paid when considering polydispersity. Indeed, this cost is significant with $T_s$ going
as $N_\text{bin}$ for $N_\text{bin} \gtrsim 10$. 
Another consideration for the volume-averaged
case is that multiple simulations are required for stochastic closure;
following figure~\ref{f:volavg_closure}~(a), at least 40 simulations are likely
required for most purposes and the dashed lines of figure~\ref{f:cost_nb}
should be ascended by this factor, accordingly. Thus, for monodisperse simulations,
the ensemble-averaging method is cheaper for all $\alpha_o$. 
For the $\alpha_o = 4 \times 10^{-5}$ bubble screen of previous sections, 
if the bubble population is considered polydisperse with $\sigma_p = 0.3$
and we accept $1\%$ relative errors in the stochastic closures of both methods, then 
the methods have nearly the same cost, with ensemble-averaging 
costing $81\%$ that of volume-averaging. Of course, smaller $\alpha_o$ 
will instead favor volume-averaging.

\begin{figure} 
	\centering
	\includegraphics[scale=1]{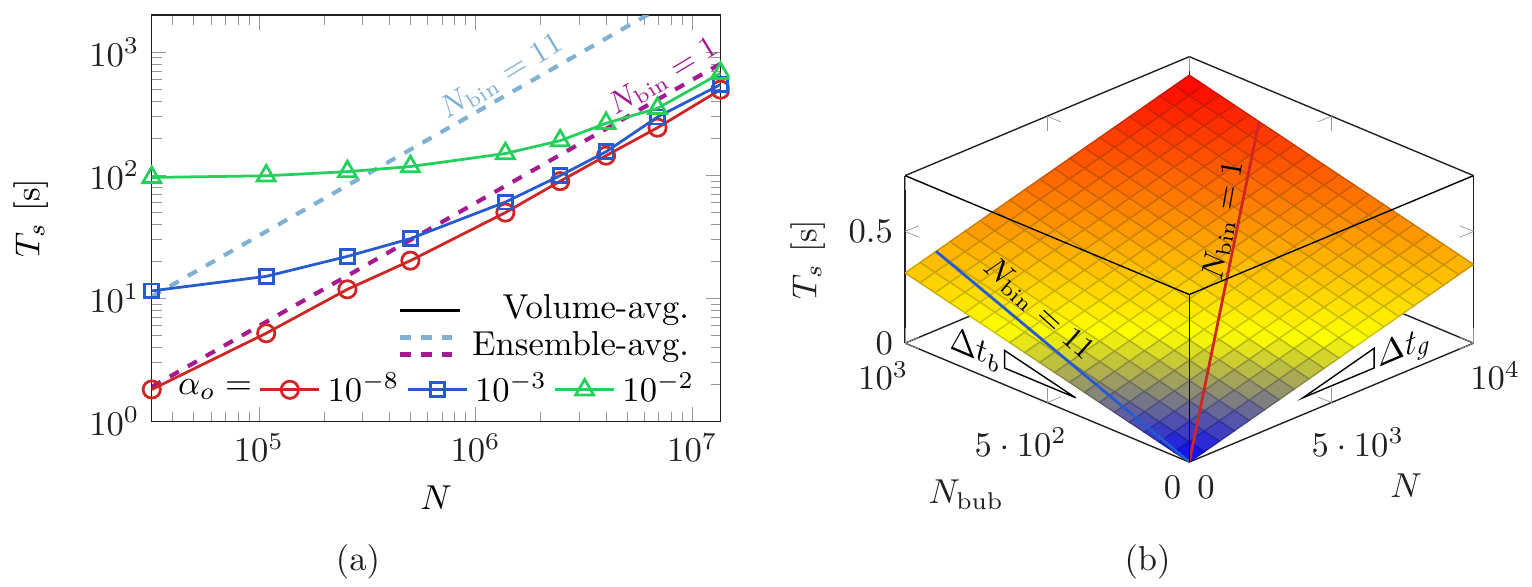}
	\caption{
	Time-step cost $T_s$ for simulations of varying spatial resolution $N = N_x N_y N_z$,
	polydispersity resolution $N_\text{bin}$, and either (a) void fraction $\alpha_o$ or (b) number of bubbles
	$N_\text{bub}$ for the volume-averaged simulations.
	}
	\label{f:cost_nx}
\end{figure}

Our final consideration is the dependence of $T_s$ on the spatial grid
resolution, and how it couples to the relative costs of polydispersity and
number of bubbles $N_\text{bub}$, where
\begin{gather}
		N_\text{bub} = \frac{3 \alpha_o }{4 \pi} \left(\frac{H}{R_o} \right)^3.
\end{gather}
Figure~\ref{f:cost_nx} (a) shows this computational cost for several example
cases. In the volume-averaged case, $T_s$ is linear with $N$ for small
$\alpha_o$, and plateaus for small $N$ if $\alpha_o$ is sufficiently large. This is
because increasing $N$ decreases the relative cost of computing the bubble
dynamics of volume-averaged simulations, as most of the time is spent
reconstructing variables and computing fluxes on the grid. For ensemble-averaged
simulations, $T_s$ increases linearly with $N$, as the bubble variables are
computed in an Eulerian framework. For small $\alpha_o = 10^{-8}$, the
volume-average cost of the bubbles is small, and the ensemble-averaged
simulations are more expensive for any $N_\text{bin}$. For larger $\alpha_o$, the
ensemble-average simulations are generally cheaper, except for cases with
sufficiently large $N_\text{bin}$.

The relative cost of simulating individual bubbles in the volume-averaged case
is shown in figure~\ref{f:cost_nx}~(b). Here, $\Delta t_b$ is the additional
time-step cost of simulating one additional bubble for fixed $N$, and $\Delta
t_g$ is this cost for one additional grid point for fixed $N_\text{bub}$. We compute
$\Delta t_b = 3.15 \times 10^{-4} \, \text{s}/N_\text{bub}$ and $\Delta t_g = 3.53 \times 10^{-5} \, \text{s}/N$, and
confirm that these values are independent within $1\%$ for varying $N$ and
$N_\text{bub}$, respectively. From these, the total time-step cost of a volume-averaged
simulation is simply $\Delta t = N_\text{bub} \Delta t_b + N \Delta t_g$, which we
confirm for independently selected cases is within $2\%$ of the actual cost. In
figure~\ref{f:cost_nx}~(b) we also label the intersection of this curve with
the cost of an ensemble-averaged simulation with polydisperse resolution $N_\text{bin}$
(which is independent of $N_\text{bub}$). Volume-averaged cases with larger $N_\text{bub}$ for
constant $N_\text{bin}$ are more expensive. For example, for $N_\text{bin} = 11$ and $N = 10^3$, 
a single ensemble-averaged simulation is cheaper than a single volume-averaged
simulation when $N_\text{bub} > 916$.

\section{Discussion and conclusions}\label{s:conclusions}

We presented a computational analysis of ensemble- and volume-averaged dilute
bubbly flow models in the context of an acoustically excited dilute bubble screen.
Results showed, for the first time, that the mixture pressure at the bubble
screen center closely matched for both the mean volume-averaged and
ensemble-averaged methods.

As a step towards assessing the relative computational cost of each method, we
focused on the cost of closing the stochastic part of the models. The volume-averaged numerical
model requires multiple, deterministic simulations of heterogeneous, randomized
dilute bubble populations to converge to the homogeneous averaged flow. We showed that
the error in the mean flow approximation decreased as $\mathcal{O}(N_\text{sim}^{-1})$,
with the associated coefficient setting the required number of simulations for
stochastic closure within a given error bound. In the case of an acoustically
excited bubble screen, error in the bubble-screen averaged pressure was about
$1 \%$ for $N_\text{sim} = 40$, and independent of the spatial resolution.
Polydisperse ensemble-averaged simulations require multiple ($N_\text{bin}$) samples of
the log-normal PDF of most-probable equilibrium bubble radii. We showed that the
error associated with undersampling this PDF decreased approximately exponentially with
increasing $N_\text{bin}$, with slower decay for larger PDF standard deviations.
Ultimately, $N_\text{bin} \gtrsim 10$ was required for faithful representation of the
polydisperse flow physics. Together, these analyses provided a framework for
computing total computational effort.

In the polydisperse case, the cost of ensemble-averaged simulations was
dominated by its stochastic closure. That is, the additional reconstructed
variables and computed fluxes on the Eulerian grid associated with the
underlying PDF were the primary time-cost of simulation for $N_\text{bin} > 5$. In such
cases, ensemble-averaged simulations were generally more expensive than their
volume-averaged counterparts. However, monodisperse simulations were generally
cheaper for the ensemble-averaged case, as only four additional equations were
added to the governing system and no individual bubble dynamics needed to be
resolved. These relative costs were complicated by the separate costs of
computing single-bubble dynamics and the Euler flow equations in the
volume-averaged case. For this, we linearly decomposed these costs such that the
relative cost of each method could be assessed for any combination of $N$,
$\alpha_o$ (or $N_\text{bub}$), and $N_\text{bin}$. For low void fraction simulations on large
spatial grids, the relative cost of computing single-bubble dynamics was small
and volume-averaged simulations were preferable. For larger void fractions on
relatively coarse meshes, the relative cost of computing the bubble-dynamic
equations was large and ensemble-average simulations were preferable.

\section*{Acknowledgements}

This work was supported by the Office
of Naval Research under grant numbers N0014-17-1-2676
and N0014-18-1-2625.

\bibliography{ref}

\end{document}